\documentclass{ws-procs9x6}

\usepackage{color}

\begin{document}

\title{FINITE TEMPERATURE CASIMIR EFFECT IN THE PRESENCE OF EXTRA DIMENSIONS }

\author{L. P. Teo}

\address{  Faculty of Engineering, University of Nottingham Malaysia Campus,\\
Semenyih, 43500, Selangor, Malaysia\\
 E-mail: LeePeng.Teo@nottingham.edu.my}

\author{K. Kirsten}

\address{Department of Mathematics, Baylor University,\\ Waco, Texas 76798-7328, USA\\
E-mail: klaus\_kirsten@baylor.edu}

\begin{abstract}
We consider the finite temperature Casimir force acting on two parallel plates in a closed cylinder with the same cross section   of arbitrary shape in the presence of extra dimensions. Dirichlet boundary conditions are imposed on one plate and fractional Neumann conditions with order between zero (Dirichlet) and one (Neumann) are imposed on the other plate. Formulas for the Casimir force show that it is always attractive for Dirichlet boundary conditions, and  is always repulsive when the fractional order is larger than $1/2$. For some fractional orders less than $1/2$, the Casimir force  can be either attractive or repulsive depending on the size of the internal manifold and temperature.
\end{abstract}


\bodymatter

\section{Introduction}\label{sec1}
  Piston configurations have received considerable interest because they allow for an unambiguous prediction of Casimir forces \cite{1}. The force depends on the boundary conditions imposed, the cross section of the piston and on properties of additional Kaluza-Klein dimensions that might be present\cite{2,3,4,5,6,7}. In this contribution we consider the same geometrical setting as in refs.\cite{4,5}, but with fractional boundary conditions imposed. This is an extension of ref. \cite{6} to the case where additional dimensions are considered.
\section{Finite temperature Casimir effect in spacetimes with extra dimensions}
 Consider a massive scalar field $\varphi(x)$ of mass $m$ in the background spacetime $M\times\mathcal{N}$, where $M$ is the  $(d+1)$-dimensional Minkowski spacetime  and $\mathcal{N}$ is a   compact $n$-dimensional internal manifold. We want to find the Casimir force acting between two parallel plates with arbitrary shape in  a  cylinder of constant simply connected cross section $\Omega$. To interpolate between Dirichlet and Neumann boundary conditions, we consider fractional Neumann conditions $$\left.\frac{\partial^{\eta_1}}{\partial x^{1\eta_1}}\varphi(x)\right|_{x^1=0}=0,\hspace{1cm} \left.\frac{\partial^{\eta_2}}{\partial x^{1\eta_2}}\varphi(x)\right|_{x^1=a}=0, \hspace{1cm}\eta_1,\eta_2\in [0,1],$$ on the plates. It can be shown that the Casimir force only depends on the difference $\eta_1-\eta_2$. Therefore without loss of generality, we fix Dirichlet boundary conditions on the left plate and impose fractional Neumann conditions of order $\eta$ on the right plate. For concreteness, on the walls of the cylinder Dirichlet boundary conditions are imposed.

Using the piston approach\cite{1}, we first consider the Casimir force acting on a piston moving freely
inside a closed cylinder $[0, L]\times \Omega \times \mathcal{N}$, and let the right end of the cylinder go to infinity, i.e., $L\rightarrow \infty$.
Let $x^1=a$ denotes the position of the piston. The finite temperature Casimir force acting on the piston is given by
\begin{align}\label{eq10_12_2}F_{\text{Cas}}^{\text{piston}}(a)=-\frac{\partial}{\partial a}\left\{E_{\text{Cas}}^{\text{cylinder}}(a)+E_{\text{Cas}}^{\text{cylinder}}(L-a)\right\},\end{align}where $E_{\text{Cas}}^{\text{cylinder}}(a)$ is the finite temperature Casimir energy inside the closed cylinder $[0, a]\times \Omega \times \mathcal{N}$; formally
\begin{equation}\label{eq10_12_1}
\begin{split}
E_{\text{Cas}}^{\text{cylinder}}(a)=\sum_{k=-\infty}^{\infty} \sum_{j=1}^{\infty}\sum_{l=0}^{\infty}\left\{\frac{1}{2}\omega_{k,j,l}  +T\log\left( 1-e^{-\omega_{k,j,l}/T}\right)\right\}.
\end{split}
\end{equation} The eigenfrequencies of the field $\varphi(x)$ which satisfies Dirichlet boundary conditions on the plate at $x^1=0$ and the wall $[0,a]\times \partial\Omega\times\mathcal{N}$ of the cylinder, and fractional Neumann conditions of order $\eta$ on the plate at $x^1=a$, are given by
$$\omega_{k,j,l} = \sqrt{\left(\frac{\pi\left(k-\frac{\eta}{2}\right)}{a}\right)^2+\omega_{\Omega,j}^2+\omega_{\mathcal{N},l}^2+m^2}
=\sqrt{\left(\frac{\pi\left(k-\frac{\eta}{2}\right)}{a}\right)^2+m_{ j,l}^2}.$$   Here, $\omega_{\Omega,j}^2$, $j=1,2,\ldots$, are the eigenvalues for the Laplace operator with Dirichlet boundary conditions on $\Omega$, $\omega_{\mathcal{N},l}^2$, $l=0,1,2,\ldots$, are the eigenvalues of the Laplace operator on $\mathcal{N}^n$, with $\omega_{\mathcal{N},0}^2=0$. For generic $\eta\in (0,1)$, $k\in \mathbb{Z}$. For $\eta=0$ or 1, $k\in \mathbb{N}$. In the following we shall  assume $\eta\in (0,1)$; for $\eta =0$ and $\eta =1$ the Casimir force will have to be divided by two.

It is worth mentioning that if one considers a cylinder of length $2a$ instead of $a$ and if one imposes twisted boundary conditions $$\varphi(x^1=2a)=e^{-i\pi \eta}\varphi(x^1=0), \quad  \varphi '(x^1=2a)=e^{-i\pi \eta}\varphi '(x^1=0)$$ for a complex massive scalar field $\varphi(x)$, one obtains the same set of eigenfrequencies.

Using the zeta regularization method to compute the Casimir energy \eqref{eq10_12_1}, we find that
\begin{align*}
&E_{\text{Cas}}^{\text{cylinder}}(a)=\Lambda_0+a\Lambda_1\\&-T\text{Re}\sum_{k=1}^{\infty}\sum_{j,l}\sum_{p=-\infty}^{\infty}\frac{e^{i\pi k\eta}}{k}\exp\left(-2ka\sqrt{m_{j,l}^2+(2\pi p T)^2}\right),
\end{align*}where $\Lambda_0$ and $\Lambda_1$ are terms that are independent of $a$. $\Lambda_0+a\Lambda_1$ does not contribute to the Casimir force \eqref{eq10_12_2} acting on the piston, and we find by taking the limit $L\rightarrow \infty$ in \eqref{eq10_12_2} that the Casimir force acting between two parallel plates is
\begin{equation}\label{eq10_12_3}F^{\parallel}_{\text{Cas}}(a)= -2T\text{Re}\;\sum_{j,l}\sum_{p=-\infty}^{\infty}\frac{e^{i\pi \eta}\sqrt{m_{j,l}^2+(2\pi p T)^2}}{\exp\left(2a\sqrt{m_{j,l}^2+(2\pi p T)^2}\right)-e^{i\pi  \eta}}.\end{equation}This expression shows that in the high temperature limit, the leading term of the Casimir force is linear in $T$ given by the sum of the terms with $p=0$, which is usually called the classical term.  For the low temperature behavior, some computations give (we use the notation $M_{k,p}^2 = k^2 a^2+(p/(2T))^2$)
\begin{eqnarray*}
F_{\text{Cas}}^{\parallel }(a)  &=& -\frac{1}{\pi  }\sum_{k=1}^{\infty}\sum_{j,l}\cos\left(\pi k\eta\right)\left\{\frac{ m_{ j,l} }{k a} K_1\left(2ka m_{j,l}\right)+2m_{j,l}^2K_0\left(2ka m_{j,l}\right)\right\}\\
&+&\frac{2 }{\pi}\sum_{k=1}^{\infty}\sum_{p=1}^{\infty}\sum_{j,l}\cos(\pi k\eta)
\frac{m_{j,l}}{M_{k,p}} \times \\
& &\hspace{2.0cm}\left\{ K_1\left(2m_{j,l} M_{k,p} \right)
-\frac{2a^2k^2m_{j,l}}{M_{k,p}} K_2\left(2m_{j,l}M_{k,p} \right)\right\}.
\end{eqnarray*}
The first two terms are the zero temperature Casimir force. The sum of the last two terms is the thermal correction. It goes to zero exponentially fast when $T\rightarrow 0$.

For a pair of Dirichlet-Dirichlet plates, i.e., $\eta=0$, \eqref{eq10_12_3} (divided by two) shows that the Casimir force is always attractive (negative).
For Dirichlet-Neumann plates, i.e., $\eta=1$, \eqref{eq10_12_3} (divided by two) shows that the Casimir force is always repulsive. In fact, this is true for all $\eta\geq 1/2$ since the function
\begin{equation}\label{eq10_12_4}
f(u)=-\text{Re}\; \left\{\frac{e^{i\pi  \eta}}{u-e^{i\pi \eta}}\right\}=\frac{1-u\cos(\pi \eta)}{(u-\cos(\pi \eta))^2+\sin^2(\pi \eta)}
\end{equation}is positive for $\eta\geq 1/2$  and $u\geq 0$.   In all these cases the force is enhanced by the presence of extra dimensions. For $\eta\in (0,1/2)$, one can show that when the plate separation is large enough, the Casimir force eventually becomes attractive.

When the plate separation $a$ is much smaller than the size $r=\text{Vol}(\Omega)^{1/(d-1)}$ of the cross section, some computations show that the leading term of the Casimir force density acting on the plates is
\begin{eqnarray}
  \frac{F_{\text{Cas}}^{\parallel}(a)}{\text{Vol}\,(\Omega)} &\sim &-\frac{4T }{2^{d}\pi^{\frac{d}{2}}}\sum_{k=1}^{\infty}\sum_{l}\sum_{p=-\infty}^{\infty}\cos(\pi k \eta)\label{eq10_12_5}\\
  & &\hspace{-1.0cm}\times \Biggl\{(d-1)\left(\frac{\sqrt{m_{l}^2+(2\pi p T)^2 }}{ka}\right)^{\frac{d}{2}}   \times K_{\frac{d}{2} }\left(2ka \sqrt{m_{l}^2+(2\pi p T)^2}\right)\nonumber\\
  & & \hspace{-1.0cm}+2\frac{\left(\sqrt{m_{l}^2+(2\pi p T)^2 }\right)^{\frac{d+2}{2}}}{\left(ka\right)^{\frac{d-2}{2}}} K_{\frac{d-2}{2} }\left(2ka \sqrt{ m_{l}^2+(2\pi p T)^2 }\right)\Biggr\},\nonumber
\end{eqnarray}where $m_l=\sqrt{\omega_{\mathcal{N},l}^2+m^2}$. In the massless case ($m=0$), the sum of the terms with $l=p=0$ should be replaced by
 $$-\frac{ (d-1)T}{2^{d-1}\pi^{\frac{d}{2}}}\frac{\Gamma\left(\frac{d}{2}\right) }{a^{d_1}}\sum_{k=1}^{\infty}\frac{\cos(\pi k\eta)}{k^{d}}.$$   A form of \eqref{eq10_12_5} more suitable to study the low temperature behavior is 
\begin{equation}\label{eq10_12_6}
\begin{split}
&  \frac{ F_{\text{Cas}}^{\parallel }(a)}{\text{Vol}\,(\Omega)}  \sim  \frac{2}{2^{d}\pi^{\frac{d_1+1}{2}}}\sum_{k=1}^{\infty}\sum_{p=-\infty}^{\infty}\sum_{ l} \cos(\pi k\eta) \left(\frac{m_l}{M_{k,p}}\right)^{\frac{d+1}{2}}\\&\times
\Biggl\{K_{\frac{d+1}{2}}\left(2m_lM_{k,p}\right)-\frac{2 k^2a^2m_l}{M_{k,p}} K_{\frac{d+3}{2}}\left(2m_lM_{k,p}\right)\Biggr\}.\end{split}
\end{equation}
When $m=0$, the sum of the terms with $l=0$ in \eqref{eq10_12_6} should be replaced by\newpage\noindent
\begin{align*}&-\frac{ d}{2^{d}\pi^{\frac{d+1}{2}}}\frac{\Gamma\left(\frac{d+1}{2}\right) }{a^{d+1}}\sum_{k=1}^{\infty}\frac{\cos(\pi k\eta)}{k^{d+1}}-\frac{2\Gamma\left(\frac{d+1}{2}\right)\zeta_R(d+1)}{\pi^{\frac{d+1}{2}}}T^{d+1}\\
&+\frac{\pi}{2^{\frac{d-2}{2}} }\frac{T^{\frac{d-2}{2}}}{a^3}\sum_{k=-\infty}^{\infty}\sum_{p=1}^{\infty}\left(k-\frac{\eta}{2}\right)^2
\left(\frac{ |k-\frac{\eta}{2}|}{pa}\right)^{\frac{d-2}{2}}K_{\frac{d-2}{2}}
\left(\frac{\pi p \left( |k-\frac{\eta}{2}|\right)}{aT}\right).
\end{align*}One can show that when $am\ll 1$, the massive correction to the Casimir force is of order $(am)^2\log(am)^2$. Therefore, \eqref{eq10_12_5} and \eqref{eq10_12_6} show that when $am\ll 1$ and $R\ll a\ll r$, where $R=\text{Vol}(\mathcal{N})^{1/n}$ is the size of the internal manifold, the nature of the Casimir force is determined by the sign of
\begin{align*}
-\sum_{k=1}^{\infty}\frac{\cos(\pi k \eta)}{k^{q}}=-B_q(\eta), \hspace{1cm} 0\leq \eta\leq 1,
\end{align*}with $q=d$ if $aT\gg 1$ and $q=d+1$ if $aT\ll 1$. For any $q$, the function $B_q(\eta), \eta\in [0,1]$ is a monotonically decreasing function and it has a unique zero $\eta_q$ in $[0,1]$. One can show that $1/3=\eta_1<\eta_2<\ldots <1/2$. Therefore, we find that when  $R\ll a \ll r$, for $\eta_{d}<\eta<\eta_{d+1}$, the Casimir force can change from attractive to repulsive when one increases the temperature. On the other hand, when the plate separation $a$ becomes much smaller than the size of the internal manifold, then the internal manifold should be treated in the same way as the large cross section. In this case, we find that the sign of the Casimir force is determined by $-B_q(\eta)$ with $q=d+n$ if $aT\gg 1$ and with $q=d+n+1$ if $aT\ll 1$. Again, we see that decreasing the distance between the plates such that it is much smaller than the size of the internal manifold can change the Casimir force from repulsive to attractive for some values of $\eta$.

\end{document}